%% file: main.tex
\documentclass[conference, final, a4paper]{IEEEtran}
\usepackage[nolist]{acronym}
\usepackage{tikz}
\usetikzlibrary{decorations.pathreplacing,calligraphy,calc}
\usepackage{pgfplots}
\usepgfplotslibrary{groupplots}
\usepackage{graphicx}
\usepackage[font=footnotesize]{caption}
\usepackage{subcaption}
\usepackage{float}
\pgfplotsset{compat=newest}
\usepackage[normalem]{ulem}
\usepackage{units}
\usepackage{amsmath, amsbsy, amssymb}
\usepackage{tikzscale}
\usepackage{color}
\usepackage{epigraph}
\usepackage{circuitikz}
\usepackage{multirow}
\usepackage{booktabs}
\usepackage[ruled]{algorithm2e}

\usepackage[group-separator={,}]{siunitx}
\definecolor{darkgreen}{rgb}{0.125,0.5,0.169}
\usetikzlibrary{shapes,arrows,fadings,chains}
\usepackage{marvosym}
\usepackage{bbm}
\usepackage{mathtools}

\tikzset{>=latex}

\input{macros.tex}

\input{acronyms.tex}
\input{corporateColours.tex}

\input{plotcyclelists.tex}

\newcommand\comjc[1]{{\textcolor{black}{{#1}}}}

\newcommand{\CLASSINPUTtoptextmargin}{25.4mm}
\newcommand{\CLASSINPUTbottomtextmargin}{19.1mm}

\newcommand*{\dittoclosing}{-- \raisebox{-0.6ex}{''} --  }

\IEEEoverridecommandlockouts

\begin{document}

\title{Optimizing Serially Concatenated Neural Codes with  Classical Decoders}

\author{\IEEEauthorblockN{Jannis Clausius, Marvin Geiselhart and Stephan ten Brink}

\IEEEauthorblockA{
 Institute of Telecommunications, University of Stuttgart, Pfaffenwaldring 47, 70659 Stuttgart, Germany \\
\{clausius,geiselhart,tenbrink\}@inue.uni-stuttgart.de\\
}

\thanks{This work is supported by the German Federal Ministry of Education and Research (BMBF) within the project Open6GHub under grant 16KISK019 and the project FunKI  under grant 16KIS1187.}
}

\maketitle
\begin{abstract}
For improving short-length codes, we demonstrate that classic decoders can also be used with real-valued, neural encoders, i.e., deep-learning based ``codeword'' sequence generators. Here, the classical decoder can be a valuable tool to gain insights into these neural codes and shed light on weaknesses. Specifically, the turbo-autoencoder is a recently developed channel coding scheme where both encoder and decoder are replaced by neural networks. We first show that the limited receptive field of \ac{CNN}-based codes enables the application of the BCJR algorithm to optimally decode them with feasible computational complexity. These \ac{MAP} component decoders then are used to form classical (iterative) turbo decoders for parallel or serially concatenated \ac{CNN} encoders, offering a close-to-\ac{ML} decoding of the learned codes. To the best of our knowledge, this is the first time that a classical decoding algorithm is applied to a non-trivial, real-valued neural code.  Furthermore, as the BCJR algorithm is fully differentiable, it is possible to train, or fine-tune, the neural encoder in an end-to-end fashion.

\end{abstract}

\acresetall

\section{Introduction}

 Over the past decades, powerful and efficient coding schemes \comjc{for long block lengths} such as turbo codes \cite{berrou1993near} and \ac{LDPC} codes \cite{GallagerLDPC} have been developed which perform near the Shannon capacity of the \ac{AWGN} channel. 
However, good encoder-decoder pairs for short block lengths (i.e., in the order of a hundred bits) which are required for \ac{URLLC} scenarios are yet to be found.

Meanwhile, deep learning based methods made their way into communication systems and channel coding in particular. 
Early works \cite{Gruber17} \cite{tandler19recurrent} demonstrated that general \acp{NN} are able to learn the decoding of classically (i.e., polar and convolutionally) coded sequences.
Models working on the Tanner graph of a code were shown to even outperform classical decoding algorithms for short codes \cite{nachmani2016learning} \cite{cammerer2022gnn}.

Soon, the question arose whether also the encoder (i.e., the code itself) can be learned as a \ac{NN}. The use of an autoencoder to jointly optimize encoder and decoder was proposed in \cite{OSheaHoydis2017introduction}. The main difference to previous methods is that the autoencoder is trained end-to-end in an unsupervised fashion.

However, na\"ive implementations of autoencoders do not scale to practical block lengths and, thus, do not exhibit competitive coding gains.
The authors of \cite{jiang19turboae} propose the turbo-autoencoder to solve these issues.
The main idea is to use two or more concatenated \ac{CNN}-based encoder networks together with iterative decoding networks --- the same principle as used in turbo codes.

\comjc{Recent research of the turbo-autoencoder involved speeding up training using Gaussian pre-training \cite{clausius21serialAE}. Others interpreted turbo autoencoder and classically decoded it based on an equivalent binary, non-linear generator polynomial \cite{Devroye2022interpreting}. In contrast, we focus on real-valued neural codes.}

Moreover, \cite{doerner2022jointautoencoder} demonstrated that the serial turbo-autoencoder can be extended to also perform joint detection, synchronization and equalization, outperforming state-of-the-art classical schemes.
Another related concept of learning encoder and decoder, using the archetype of polar codes (rather than turbo codes), was proposed in \cite{kocodes21}.

\begin{table}[t]
	\caption{\footnotesize Selected related work on classical and deep-learning based channel coding schemes }
	\label{tab:intro}
	\centering
	\begin{tabular}{p{17mm}|p{21.8mm}|p{36mm}} 
		& Classical Decoding & Deep Learning based Decoding \\ %
		\hline
		Classical \mbox{Encoding} & Turbo codes \cite{berrou1993near}\newline LDPC codes \cite{GallagerLDPC} & Dense Neueral Networks \cite{Gruber17}\newline RNN-based conv. decoding \cite{tandler19recurrent}\newline Weighted belief propagation \cite{nachmani2016learning}\newline Graph Neural Networks \cite{cammerer2022gnn}\\
		\hline
		\comjc{Deep Neural Encoder} & \textbf{This work} & Autoencoder \cite{OSheaHoydis2017introduction}\newline Turbo Autoencoder \cite{jiang19turboae} \newline K.O. codes \cite{kocodes21}
	\end{tabular}
\vspace{-0.5cm}
\end{table}

Table \ref{tab:intro} arranges these prior works in two dimensions: Whether classical (i.e., engineered) or deep \comjc{\acp{NN}} are used for the encoder and decoder, respectively. The evolution outlined above moved from classical systems to first making the decoder trainable, and then learning the encoder as well. A natural question to ask is whether it is also possible to use classical decoders with neural encoders. 
As classical decoders are better understood, this can serve as a valuable tool to analyze and fine-tune neural codes.

The main contributions of this paper are as follows:

\begin{itemize}
   \item \comjc{We derive a low-complexity method for analyzing the effective memory of the \ac{CNN} encoders and show that it is shorter than what their receptive field would allow.}
    \item We present how the \ac{BCJR} algorithm can be applied to decode learned, real-valued \ac{CNN} codes with quasi-\ac{MAP} performance. 
    The \ac{BCJR}-based turbo decoder is then applied to decode trained \acp{SCNC}.
    \item We show that using the \ac{BCJR}-based decoder, neural encoders can be fine-tuned or even trained from scratch to obtain improved performance.
\end{itemize}

\section{Preliminaries}
\label{sec:sys}

\subsection{Convolutional Neural Networks}
\begin{figure}[t]
	\centering
	\resizebox{0.8\columnwidth}{!}{\input{tikz/cnn}}
	\caption{\footnotesize Visualization of the receptive field of a \ac{CNN} with kernel size \mbox{$\kappa=5$}. The positional axis is arranged vertically, the different feature maps are indicated in the depth dimension. }
	\label{fig:cnn_structure}
	\vspace{-0.5cm}
\end{figure}
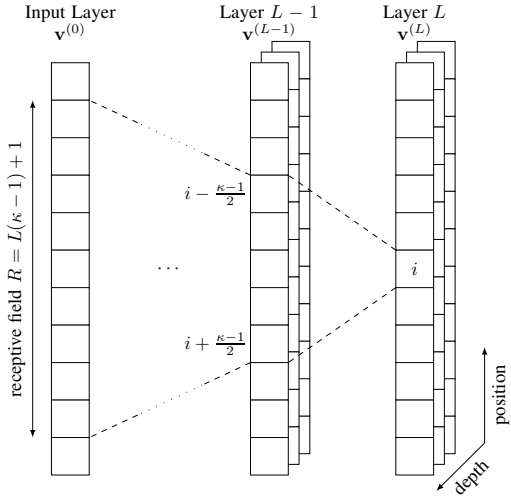

A commonly used class of feed-forward neural networks are \acp{CNN}. 
Each layer can be mathematically described by a convolution operation of the input and trainable filter kernels, followed by a non-linear activation function. 
The \acp{CNN} used in the turbo autoencoder use one-dimensional (1D) convolutions over the \textit{positional} axis. 
Each layer can have a depth of multiple such one-dimensional \textit{feature maps} and has independent trainable filter kernels of size $\kappa$ for each combination of input and output feature maps.
As illustrated in Fig.~\ref{fig:cnn_structure}, the output of any layer only depends on $\kappa$ elements of the previous layer. Hence, after $L$ convolutional layers with kernel size $\kappa$, the number of input symbols affecting  each output symbol is limited to $R=L\cdot(\kappa-1)+1$. This number is called the \textit{receptive field} of the \ac{CNN}.

\subsection{Serial Turbo Autoencoder based System}
\label{sec:ae}

\begin{figure}[t]
	\centering
	\resizebox{0.48\textwidth}{!}{\input{tikz/serial_encoder}}
	\caption{ \footnotesize Serially concatenated CNN-based encoder structure of the transmitter learning a real-valued \acf{SCNC}.}

	\label{fig:serial_encoder}
	\vspace{-0.35cm}
\end{figure}
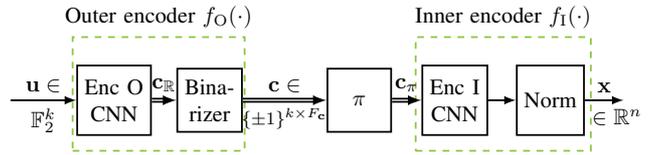

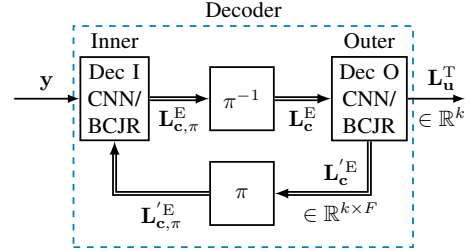
\begin{figure}[t]
	\centering
	\resizebox{0.35\textwidth}{!}{\input{tikz/serial_decoder}}
	\caption{\footnotesize Serial CNN- or BCJR-based turbo decoder structure.
	}
    \vspace{-0.25cm}
	\label{fig:serial_decoder}
\end{figure}

General-purpose \acp{DNN} or \acp{CNN} are not able to learn codes that are comparable to state-of-the-art (binary) channel codes due to the \textit{curse of dimensionality} \cite{jiang19turboae,ye19circularAE}. To overcome this problem and learn competitive, neural channel codes for messages $\uv$ longer than $k>16$ bits, the serial turbo-autoencoder \cite{jiang19turboae,clausius21serialAE} has been proposed.
The serial turbo-autoencoder is an autoencoder that employs the structure of serially concatenated codes with an iterative decoder.
The architecture is displayed in Fig.~\ref{fig:serial_encoder} for the encoder and in Fig.~\ref{fig:serial_decoder} for the decoder.
While the overall structure of this end-to-end trainable system is carefully chosen, the constituent components are general-purpose \acp{CNN}. In addition to the proposed  turbo-autoencoders in \cite{jiang19turboae,clausius21serialAE}, we employ the advancements of circular convolutions from \cite{yildiz21interleaver,ye19circularAE} and the linear interleaver from \cite{yildiz21interleaver}. Note, that the circular padding in the encoder can be interpreted as tail-biting encoding.

Fig.~\ref{fig:serial_encoder} shows the encoder of the serial turbo-autoencoder.
It consists of a cascade of an outer encoder $f_\mathrm{O}(\cdot)$, an interleaver and an inner encoder $f_\mathrm{I}(\cdot)$.
The hyperparameters that are used for all \acp{CNN} are shown in Tab.~\ref{tab:hyperparameter}. 
The \ac{CNN} of the outer encoder sees the uncoded message $\uv \in \left\{-1,1 \right\}^{k}$ and encodes it into a real-valued sequence $\cv_{\RR}^{k \times F_{\cv}}$, where $F_{\cv}$ is a hyperparameter denoting the depth of the outputs (illustrated  by  double  arrows  in  Fig.~\ref{fig:serial_encoder}).
Subsequently, the real-valued, coded sequence $\cv_{\RR}$ is binarized, so  $\cv = \operatorname{sign}\left(\cv_{\RR}\right)$. Note, as the derivative $\frac{\partial \operatorname{sign}(x)}{\partial x}=0$ almost everywhere, we replace the derivative for back-propagation with a linear function with saturation effects

\begin{equation}
\frac{\partial \operatorname{sign}(x)}{\partial x} = \begin{cases}
    1 &\text{for $|x|<1$}\\
    0 &\text{else.}
    \end{cases}
\end{equation}
This is known as the saturated \ac{STE} \cite{BengioLC13}. 
In a next step, the coded sequence $\cv$ with $k$ \emph{positional} entries and $F_\mathrm{\cv}$ \emph{depth} entries is interleaved over the positional dimension.
Finally, the inner encoder (Enc I)\comjc{, which acts also as modulator,} computes the $n$ symbols $\xv^{k\times F_\xv}$ which then are normalized and transmitted. 
\comjc{Note, that the hyperparameter $F_\xv$ defines the rate $r=\nicefrac{k}{n}=\nicefrac{1}{F_\xv}$ of the \ac{SCNC}.}

The iterative structure for decoding serially concatenated codes is shown in Fig.~\ref{fig:serial_decoder}.
\comjc{The inner decoder (Dec I) sees the channel observations $\yv =\xv +\nv$ with $\nv\sim\mathcal{N}(0,\sigma^2)$ and the a priori information from the outer decoder (Dec O). }
In the first iteration the a priori information is initialized as  $\Lm_{\cv,\pi}^{'\mathrm{E}}=\mathbf{0}$, where the subscript, i.e., $\cv$, denotes the bits which the information is about. 
Then the extrinsic information  of the inner decoder is interleaved and forwarded to the outer decoder. The outer decoder outputs an estimate $\Lm_{\uv}^\mathrm{T}$ of the uncoded bits $\uv$ and the extrinsic information $\Lm_{\cv}^{'\mathrm{E}}$, which is the a priori information for the inner decoder. 

The serial turbo-autoencoder can be trained in an end-to-end fashion as all components are differentiable. 
We employ the same training methods as in \cite{jiang19turboae}.
The hyperparameters for the training are shown in Tab.~\ref{tab:hyperparameter}. 
Note, we use a random interleaver for training  and a linear interleaver during inference.
As loss function we chose the \ac{BCE} function.
It is worth mentioning that the learning schedule for the encoder and decoder is alternating. That means, in a loop we calculate $T_\mathrm{TX}$ \ac{SGD} updates for the whole encoder while the decoder weights are fixed, and in a next step, we calculate $T_\mathrm{RX}$ \ac{SGD} updates for the whole decoder while the encoder weights are fixed.

Similar to \cite{clausius21serialAE} and \cite{doerner2022jointautoencoder} we observe a significant amount of single-bit errors after decoding.
This implies that the bit errors in the decoder are loosely correlated. 
As a result, the \ac{BLER} can be drastically improved by 
an outer single-parity-check code
and flipping the bit with the least reliability in case of a not matching parity.
We consider the associated rate loss in form of a rate shift of $10\log_{10}(\nicefrac{k-1}{k})$~dB.

\begin{table}
	\caption{\footnotesize Hyperparameters for architecture and training the serial turbo-autoencoder}
	\label{tab:hyperparameter}
	\centering
	\begin{tabular}{l|l}
		Parameter& Value \\
		\hline
		CNN filters & 100\\
		CNN kernel & 5\\
		Layers per CNN & 5 \\
		Receiver iters. & 6 \\
		$F_c$, $F_\xv$ & 10, 2 \\
		$(k,n)$ & $(64,128)$\\
		Padding & Circular
		
	\end{tabular}
	\begin{tabular}{l|l}
		Parameter& Value \\
		\hline
		Loss & BCE \\
		$T_{\mathrm{TX}}$, $T_{\mathrm{RX}}$  & $100$, $500$  \\
		Batch size & $500-4000$  \\
		Optimizer & ADAM \\
		Learning rate & $10^{-4}-10^{-6}$  \\
		Encoder SNR & $4.0~\mathrm{dB}$  \\
		Decoder SNR & $0.5-4.0~\mathrm{dB}$ \\

	\end{tabular}
	\vspace{-0.5cm}
\end{table}

\subsection{Overview of the BCJR Algorithm}
\label{sec:bcjr}
This subsection shortly introduces the log-domain \ac{BCJR} algorithm \cite{ryan2009channelcodes} for decoding in a serial turbo decoder.
The \ac{BCJR} algorithm is an optimal decoder in the sense of the bit-wise \ac{MAP} criterion.
This decoding criterion is formulated by $\arg \max_{u_i} p(u_i |\yv)$, where the bit $u_i$ is estimated based on the noisy sequence $\yv$.
\comjc{While the decoding of every component code is optimal, the turbo decoder needs a sufficient amount of iterations until it converges close to the optimal performance of the concatenated code. }
The algorithm is based on the following equations:
\begin{align}
    \gamma_i(s',s) &= \log( p(u_i)p(y_i|u_i)) \label{eq:gamma}\\
    \alpha_i(s) &= \underset{s'}{\max}^* \left[\alpha_{l-1}(s')+\gamma_i(s',s) \right]\label{eq:alpha}\\
      \beta_{i-1}(s') &= \underset{s}{\max}^* \left[\beta_{i}(s)+\gamma_i(s',s) \right]\label{eq:beta}\\
      \mathrm{max}^*(x,y) &\triangleq \log(\mathrm{e}^x+\mathrm{e}^y) \approx \max(x,y)\label{eq:max}
\end{align}

Equation (\ref{eq:gamma}) is a branch metric that needs to be calculated for every \comjc{state }transition \comjc{from $s'$ to $s$} where $u_i$ and $y_i$ are the $i$th bit and observation, respectively, with $i=0,\dots,N-1$.
The state transitions $(s',s)$ are defined by the encoder.
The recursive equations (\ref{eq:alpha}) and (\ref{eq:beta}) are path metrics that traverse the trellis in a forward and a backward manner, respectively. %
Whether we use equation (\ref{eq:max}) or the given approximation coincides with the labeling \emph{log-MAP} and \emph{max-log} respectively.
The performance loss due to the approximation is usually omittable, if a suitable damping factor is applied after every \ac{BCJR} decoder.
Once the branch and path metrics are calculated, a final soft decision can be made according to
\begin{multline}
    L_{\uv,i}= \underset{U^1}{\max}^* \left[ \alpha_{l-1}(s') +\gamma_i(s',s) + \beta_{i}(s) \right] \\
    - \underset{U^{-1}}{\max}^* \left[ \alpha_{l-1}(s') +\gamma_i(s',s) + \beta_{i}(s) \right]
\end{multline}
where $U^{1}$ and $U^{-1}$ are the sets that contain every transition that is associated with $u_i=1$ and $u_i=-1$, respectively.
Note that the choice of $U^{1}$ and $U^{-1}$ determines whether the output is regarding the coded or the uncoded bits.

In the case of serial turbo decoding as depicted in Fig.~\ref{fig:serial_decoder}, the inner decoder sees the channel observation $\yv$ and the a priori information  $L_{\cv,i}'^{\mathrm{E}} $ of the outer decoder. Therefore, (\ref{eq:gamma}) is calculated as
\begin{equation}\label{eq:gamma_inner}
    \gamma_i(s',s) = \frac{1}{2} L_{\cv,i}'^{\mathrm{E}} \cdot c_{s',s} - \frac{|y_i-x_{s',s}|^2}{2\sigma^2 }
\end{equation}
where \comjc{$c_{s',s}$ is the true coded bit for the transition $(s',s)$ and} $\sigma^2$ is the noise power.
In contrast, the outer decoder only receives the a priori information from the inner decoder, and thus
\begin{equation}\label{eq:gamma_out}
    \gamma_i(s',s) = \frac{1}{2} L_{\cv,i}^{\mathrm{E}} \cdot c_{s',s}.
\end{equation}
As all presented equations are differentiable we can calculate a gradient though the \ac{BCJR} decoders and, therefore, also the (unrolled) turbo decoder.
This enables the \ac{BCJR} algorithm to be used in the turbo decoder structure not only for decoding, but also training  in an end-to-end fashion.

A last comment needs to be made in this section regarding the tail-biting (circular padding) structure of the encoder.
The structure determines that the state at the first position $i=0$ equals the state after the last position $i=N$.
Thus, we need to set the initial path metrics $\alpha$ and $\beta$  accordingly.
However, the initial state is unknown. 
One way to circumvent this problem is to consider the wrap-around trellis \cite{anderson98wrapBCJR}.
The initialization of the path metrics begins $w$ positions before the first position at index $i_\mathrm{start}=-w \mod N$ and $i_\mathrm{end}=N+w \mod N $ for $\alpha$ and $\beta$, respectively.
In other words, the sequence $\yv'=\left[ y_{-w\mod N},y_{-w+1\mod N},...,y_{N+w\mod N} \right]$ is decoded instead of $\yv=\left[ y_{0},y_{1},...,y_{N-1} \right]$.

\section{BCJR decoding of Neural Codes}
\label{sec:neural_bcjr}

In this section we derive how to decode a learned \ac{SCNC} with a classical decoder, namely the \ac{BCJR} decoder. Note that for this purpose, we assume that the \acf{SCNC} has been already trained in a turbo-autoencoder, as shown, e.g., in \cite{clausius21serialAE}.

\subsection{Estimating the Effective Memory of the Neural Codes}
This subsection shows how to estimate the effective memory of a \ac{CNN} from the gradient and that the gradient is, indeed, a useful indicator.
We define that the \comjc{effective} memory of the \ac{CNN} consists of \emph{those} input positions inside the receptive field of each output \emph{that} significantly contribute to the output value.
Whether decoding of \ac{CNN}-based codes is feasible or not using the \ac{BCJR} algorithm depends on the memory size of the \acp{CNN} which exponentially relates to the number of states in the trellis.
As each \ac{BCJR} decoder computes the likelihood of every possible symbol (inner decoder) or bit (outer decoder) of the respective encoder at every position, we need to pre-calculate the possible symbols and bits for the state transitions. 
The number of possible symbols for an encoder \ac{CNN} is $M=\left| \mathcal{F} \right|^m$, where $\left| \mathcal{F} \right|$ is the number of elements of the set of the input to the \ac{CNN} and $m$ is the effective size of the memory of the \ac{CNN}.
Note that the filters in the \acp{CNN} are not causal and, thus, the memory does not only include past inputs but also future inputs.
The memory of a \ac{CNN} is upper bounded by the receptive field that is determined by the \ac{CNN} structure. 
In this case the receptive field is $R=21$ which could result in up to $M=\left| \mathbb{F}_2 \right|^{R}\approx2\cdot 10^7$ symbols for the binary field $\mathbb{F}_2$.
The usage of the binarizer ensures that all inputs to the \acp{CNN} in the encoder are binary.

As \ac{BCJR} decoding with $2^{R-1}$ states is not feasible, positions with low contribution to the output can be pruned. We propose an easy way to measure the contribution by calculating the average energy of the gradient
\begin{align}\label{eq:grad}
    e_{\partial, \vv}(j) &=  \operatorname{E}\left[ \left| \frac{\partial f(\vv)}{\partial v_j} \right|^2 \right]\\
    &\approx \frac{1}{N_\mathrm{B}}\sum_{i=0}^{N_\mathrm{B}} \left| \frac{\partial f(\vv_i)}{\partial v_{i,j}} \right|^2
\end{align}
where $\vv_i$ is the $i$th \comjc{sample that is fed} to the neural network $f(\cdot)$ from a batch of size $N_\mathrm{B}$.
We interpret the expected gradient (change of the output by changing the input) as the exhibited \comjc{expected }error of omitting an input position.
The gradient is usually a by-product of any deep learning framework, and therefore Eq.~(\ref{eq:grad}) is readily available. %
In our case it is sufficient to consider a single output, as the weight-sharing property of the \ac{CNN} and the circular padding ensure that each output exhibits the same dependencies to its relative input positions.
In a next step, we compare the energy of the gradients for each encoder to a threshold $t$ to obtain the effective size $m$ of the memory. If the energy is larger than the threshold, then we consider this position to contribute significantly to an output and, thus, is part of the memory. We denote the set of positions from the first to the last contributing position by $\mathcal{J}$ with $|\mathcal{J}|=m+1$ and all positions outside $\mathcal{J}$ by $\bar{\mathcal{J}}$.

Since the gradient only describes the behavior around the inputs $\pm1$, we need to verify that a bit flip from $\pm1$ to $\mp1$, respectively, exhibits a similar behavior.
For this, we calculate the expected energy of the difference quotient per bit position for a bit flip \cite{Devroye2022interpreting} as
\begin{align}\label{eq:diff}
    e_{\Delta,\vv}(j) &= \operatorname{E}\left[  \frac{\left|f(\vv)-f(\vv\odot(1-2\ev_j))\right|^2}{|| 2\ev_j ||^2} \right]\\ %
    &= \frac{1}{4\cdot2^{R}}\sum_{\vv\in\{\pm1\}^R} \left|f(\vv)-f(\vv\odot(1-2\ev_j))\right|^2
\end{align}
where $\ev_j$ denotes the unit vector which is 1 exactly in position $j$ and $\odot$ denotes element-wise multiplication.
Note, Eq.~(\ref{eq:diff}) corresponds to the actual expected error of omitting an input position.
In the following analysis of the encoder \acp{CNN} we empirically verify that Eq.~(\ref{eq:grad}) is, indeed, an approximation of Eq.~(\ref{eq:diff}) up to a scaling factor.

\subsection{Outer Neural Encoder with Binary Output}

\begin{figure}[t]
	\centering
	\resizebox{0.4\textwidth}{!}{\input{tikz/effective_encoder.tikz}}
	\caption{ \footnotesize Equivalent encoder block diagram. The outer encoder learns two output streams. Splitting the inputs to the inner encoder reduces the size of the memory $m_\xv^{(0)}$ and $m_\xv^{(1)}$, respectively, to allow BCJR decoding at feasible complexity.}

	\label{fig:effective_encoder}
	\vspace{-0.25cm}
\end{figure}
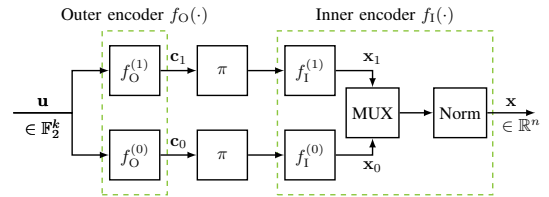

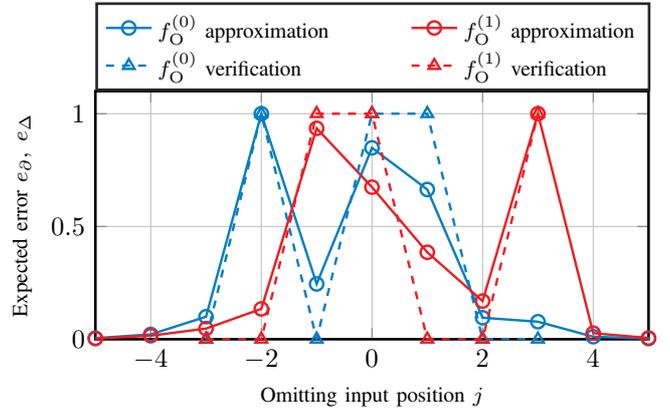
\begin{figure}[t]
	\centering
	\input{tikz/gradient_energy_outer.tikz}
    \caption{\footnotesize Estimating the effective memory through the normalized expected error ($e_{\partial}$ approximation, $e_{\Delta}$ verification) of omitting input $u_j$ in relation to each output for the outer encoders $f_\mathrm{O}(\cdot)$. \comjc{Note, the linear interpolation between points is only shown for visualization purposes.}
	}

	\label{fig:gradient_outer}
	\vspace{-0.5cm}
\end{figure}

The outer encoder implements the mapping $f_{\mathrm{O}}:\mathbb{F}_2^k \to \mathbb{F}_2^{k\times F_c} $.
This means in general the \ac{CNN} can learn an encoding with a coding rate as low as $\nicefrac{1}{F_c}$, here $F_c=10$.
We observed, however, that only two out of the $F_c$ depth dimensions of the output vary dependent on the input, while the remaining eight depth dimensions stay fixed at either $1$ or $-1$ for all possible inputs. Therefore, a rate of $\nicefrac{1}{2}$ is learned which corresponds to the overall code rate of the concatenated neural encoders.
We label the two dimensions with varying output $\cv^{(0)} = f_{\mathrm{O}}^{(0)}(\uv)$ and $\cv^{(1)} = f_{\mathrm{O}}^{(1)}(\uv)$.
This is illustrated in Fig.~\ref{fig:effective_encoder}.
It is remarkable to realize that the \ac{CNN} learns this on its own without any injection of expert knowledge.
Moreover, this observation aligns with the derivation from \cite{ashikhmin04extrinsic} that an outer code with a rate larger than the overall rate of a serial concatenated code is sub-optimal for iterative decoding.

\comjc{Note, Eq.~(\ref{eq:diff}) captures the effects of the binarizer, while Eq.~(\ref{eq:grad}) does not, due to the \ac{STE}. }
The obtained plot for $\cv_0$ and $\cv_1$ is shown in Fig.~\ref{fig:gradient_outer}.
First of all, we want to mention that we expect a deviation of the approximation Eq.~(\ref{eq:grad}) as it does not capture the influence of the binarizer.
Indeed, we observe a difference, albeit the general behavior matches.
More interesting are the obtained values for $e_{\Delta}$ which are either $1$ or $0$.
This means that a bit flip in the respective input position either leads to a bit flip in all cases or in none.
As a consequence, the outer encoder is equivalent to the linear generator polynomials $G_0(Z)=Z^{-2}+Z^0+Z^{1}$ and $G_1(Z)=Z^{-1}+Z^0+Z^{3}$.
This coincides with the reasoning from \cite[p.~320]{ryan2009channelcodes} that the outer codes of serially concatenated turbo codes may be non-recursive and linear, while the inner codes have to be recursive or non-linear.
The implication for decoding is that we can either run two \ac{BCJR} decoders with memories $m_{\cv}^{(0)} = 3$, $m_{\cv}^{(1)} = 4$ and aggregate the bit estimates or run one \ac{BCJR} decoder with memory $m_{\cv}=5$.
The first option can be chosen to save complexity. 
With this knowledge, we can pre-compute the output bits $c_{s',s}$ for each of the $2^{m_{\cv}+1}=32$ state transitions $(\mathrm{s',s})$.
Finally, we can plug $c_\mathrm{s',s}$ into the Eq.~(\ref{eq:gamma_out}) and decode the outer neural code.

\subsection{Inner Neural Encoder with Real-valued Output}
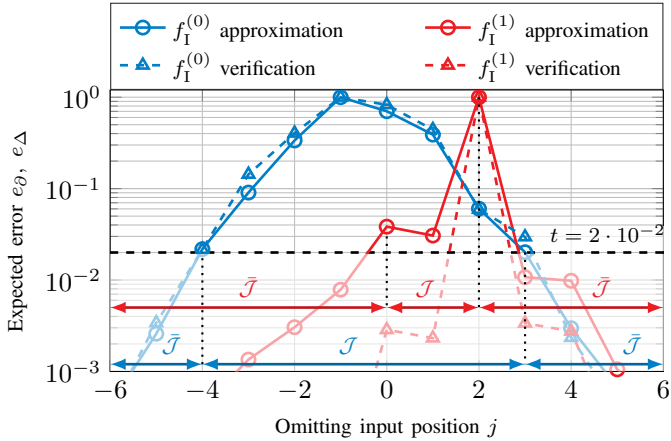
\begin{figure}[t]
	\centering
	\input{tikz/gradient_energy.tikz}
    \vspace{-0.5cm}
	\caption{\footnotesize Estimating the effective memory through the normalized expected error ($e_{\partial}$ approximation, $e_{\Delta}$ verification) of omitting input $c_j$ in relation to each output for the inner encoders $f_\mathrm{I}(\cdot)$. Positions with an error exceeding the threshold belong to the set of contributing positions $\mathcal{J}$.
}
    \vspace{-0.5cm}
	\label{fig:gradient}
\end{figure}

As we established in the last section that the coded bits $\cv$ have essentially dimensions $k\times 2$, the inner encoder learns the mapping $\mathbb{F}_2^{k\times 2}\to \mathbb{R}^{k\times 2} $.
A problem is that the second depth dimension of the input roughly doubles the memory. 
To circumvent this problem, we first encode $\cv$ with $\cv^{(1)}=\mathbf{1}$ and then encode $\cv$ with $\cv^{(0)}=\mathbf{1}$. 
This is illustrated in Fig.~\ref{fig:effective_encoder}.
Subsequently, we concatenate the first depth dimension of the first output with the second depth dimension of the second output to form output symbols $\xv =\left[\xv^{(0)},\xv^{(1)} \right]$.
Effectively, the inner encoder consists of two functions $\xv^{(0)} = f_{\mathrm{I}}^{(0)}(\cv_{\pi}^{(0)})$ and $\xv^{(1)} = f_{\mathrm{I}}^{(1)}(\cv_{\pi}^{(1)})$.
\comjc{Fig.~\ref{fig:gradient} shows Eq.~(\ref{eq:grad}) and Eq.~(\ref{eq:diff}) for both encoding functions.}
We observe that the approximation by the gradient closely follows the actual difference.
Next we can set a threshold $t$ to obtain the memory.
We found that a threshold of $t=2\cdot10^{-2}$ does not degrade the final \ac{BLER}.
This leads to memory size $m_{\xv}^{(0)}=7$ for $f_{\mathrm{I}}^{(0)}(\cdot)$ and $m_{\xv}^{(1)}=2$ for $f_{\mathrm{I}}^{(1)}(\cdot)$.
However, the verification error $e_{\Delta}$ of $f_{\mathrm{I}}^{(1)}$  indicates  that actually an encoding without significant memory is learned. Thus, the learned encoding function $f_{\mathrm{I}}^{(1)}$ can be interpreted as a scaled \ac{BPSK}.

For the calculation of $x_{s',s}$ in Eq.~(\ref{eq:gamma_inner}), we average $x=f(\cv)$ with $\cv = \left[ c_{-\nicefrac{R}{2}},...,c_{\nicefrac{R}{2}} \right]$ over all possible values of $c_j$ in positions $j\in\bar{\mathcal{J}}$ and $c_j$ fixed for all $j\in\mathcal{J}$. Note that each transition $(s',s)$ corresponds to exactly one value of the sub-vector $\cv_{\mathcal{J}}$ that contains $c_j \forall j \in \mathcal{J}$. This is done for both $\cv^{(0)}$ and $\cv^{(1)}$.

\vspace{-0.15cm}
\section{Results}
\label{sec:results}

\subsection{Decoding Performance}

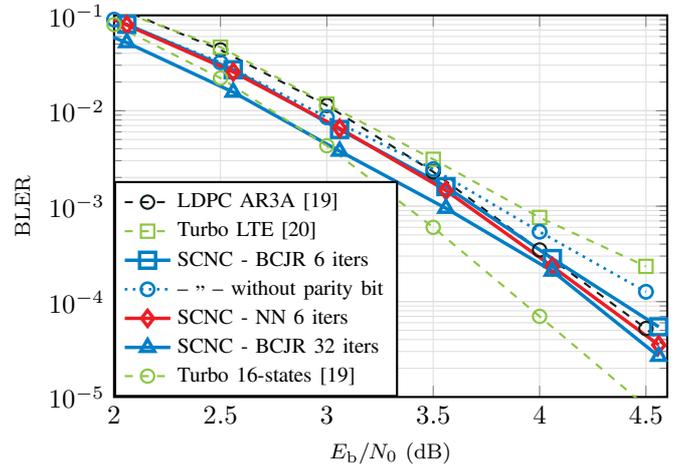
\begin{figure}[t]
	\centering
	\vspace{-0.25cm}
	\input{tikz/bler_graph.tikz}
	\vspace{-0.5cm}
	\caption{\footnotesize Simulated BLER over an AWGN channel ($k=64$, $n=128$) for the \acf{SCNC} with a \ac{BCJR}-based and a \ac{NN}-based turbo decoder, respectively.}
	\label{fig:bler}
	\vspace{-0.25cm}
\end{figure}

As performance metric we choose the \ac{BLER} over the \ac{BER}, as it is more relevant in systems with short block lengths.
In our transmission model we consider an \ac{AWGN} channel.
 Further, we aim to transmit $k=64$ bits over $n=128$ channel uses, thus $r=0.5$.
 The threshold for memory estimation is set to $t=2\cdot 10^{-2}$ and we extend the wrap-around \ac{BCJR} by $w=16$ positions in both directions.
 Fig.~\ref{fig:bler} shows the performance of the log-MAP \ac{BCJR}-based turbo decoder for the \ac{SCNC} with 6 iterations (\ref{plot:bcjr_delta11_it6_tb16}) and 32 iterations(\ref{plot:bcjr_delta11_tb32_it32_map})  respectively.
 Additionally, a curve without a parity bit and bit flip (\ref{plot:bcjr_delta11_it6_tb16_no_bf}) is displayed, as described in section~\ref{sec:ae}.
Recall that the number of states for the outer \ac{BCJR} is set to $2^{m_\cv} = 32$ and for the inner \ac{BCJR} to $2^{m_{\xv}^{(0)}}=128$ and $2^{m_{\xv}^{(1)}}=2$.
 For comparison, the same \ac{SCNC}  is shown with the matching \ac{CNN}-based turbo decoder (\ref{plot:stae_delta11}). 
 Additionally, several baselines from \cite{liva2016code} are shown and an additional LTE  turbo code (\ref{plot:turbo_lte_maxlog_64_128}) \cite{3gpp2007standard} with 8 iterations of damped max-log decoding.
We observe that the \ac{BCJR} with 6 decoder iterations (\ref{plot:bcjr_delta11_it6_tb16}) shows slightly worse performance than the \ac{CNN} decoder (\ref{plot:stae_delta11}). 
However, 32 \ac{BCJR} iterations (\ref{plot:bcjr_delta11_tb32_it32_map}) yield a significant gain for lower \acp{SNR}, but a omittable gain for higher \acp{SNR}. 
Thus, we conclude that the \ac{NN}-based decoder operates close to its optimal performance for a high \ac{SNR}.

\subsection{Insights about the Turbo-Autoencoder}

From the decoding of the \ac{SCNC} with a classical decoder, we obtained insights regarding the turbo-autoencoder. 
To summarize the observations so far,
Section~\ref{sec:neural_bcjr} showed that the outer encoder learned a linear encoding and one of the inner encoder exhibits systematic behavior.
The decoding results show that the learned \ac{NN}-based decoder slightly outperforms the log-MAP \ac{BCJR}-based decoder. 
Thus, the \ac{NN}-based decoder displays close to optimal performance for the \ac{SCNC}.
This sets the spotlight on the encoder for potential improvements in performance.
For both decoders we noticed a significant amount of single erroneous bits in a block error.
For a \ac{SNR} of $4.5$~dB the percentage of block errors with a single erroneous bit to all block errors is $71.5\%$ and $66.0\%$ for the \ac{BCJR}-based and \ac{NN}-based turbo decoder respectively. 
Conclusively, the minimal distance between code words seems to be small.
Thus, the \ac{BLER} is sub-optimal without an outer error-correcting code, which improves the minimum distance.
A major factor for this behavior might be the \ac{BCE} loss function, as it is known to minimize the \ac{BER} and not the \ac{BLER} \cite{cammerer2019tcom}.

\subsection{Training with BCJR decoders}
\begin{figure}[t]
	\centering
	\vspace{-0.25cm}
	\input{tikz/bler_train_graph.tikz}
	\vspace{-0.35cm}
	\caption{\footnotesize Simulated BLER over an AWGN channel ($k=64$, $n=64$). Showcasing the training of \acf{SCNC} for a \ac{BCJR}-based decoder with reduced memory $m_\xv^{(0)}=7\to 3$ and reduced iterations $6\to2$.}
	\label{fig:bler_train}
	\vspace{-0.45cm}
\end{figure}
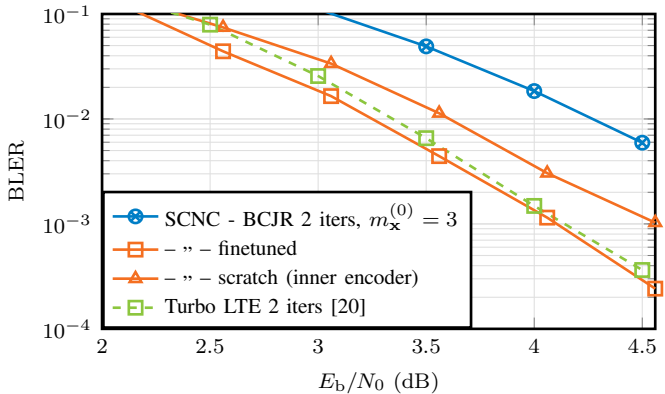

As all equations for the \ac{BCJR}-based turbo decoder are differentiable, we can train a neural encoder with a classical decoder. 
Here, we want to showcase  this possibility for a reduction of the size of the memory $m_\xv^{(0)}=7\to 3$ and the number of decoder iterations $6\to 2$. 
The performance for this case for the not fine-tuned \ac{SCNC} (\ref{plot:bcjr_delta11_it2_tb16_logmap_m5_m3_m1}) is displayed in Fig.~\ref{fig:bler_train}.
The fine-tuned \ac{SCNC} with the same decoder (\ref{plot:bcjr_delta11_it2_tb16_logmap_m5_m3_m1_finetuned}) shows significant gain. 
Also it is possible to learn an encoder from scratch, i.e., the inner encoder (\ref{plot:bcjr_delta11_it2_tb16_logmap_m5_m4_m4_scratch}).
For reference, we plot the LTE turbo code with 2 iterations of damped max-log decoding (\ref{plot:turbo_lte_logmap_64_128_2iters}).
The hyperparameters for fine-tuning are listed in Tab.~\ref{tab:hyperparameter_fine}.
We found that a high batch size and a small learning rate are essential.
Furthermore, the log-MAP \ac{BCJR} lead to a smoother training with better performance than the max-log approximation.

\begin{table}
	\caption{\footnotesize Hyperparameters for training and fine-tuning}
	\label{tab:hyperparameter_fine}
	\centering
	\begin{tabular}{l|l}
		Parameter& Value \\
		\hline
		Batch size & 1000\\
		Learning Rate & $10^{-5}$\\
		SNR & 4 \\

	\end{tabular}
	\begin{tabular}{l|l}
		Parameter& Value \\
		\hline
		Loss & BCE \\
		Receiver iters. & 2 \\
        SGD updates & 1000

	\end{tabular}
	\vspace{-0.25cm}
\end{table}

\vspace{-0.35cm}
\section{Conclusion}
\label{sec:conclusion}
In this paper, we demonstrated that the \ac{BCJR} decoding algorithm can be applied to \ac{CNN}-based encoders. To mitigate computational complexity, we showed that only a fraction of the potentially learned memory in the encoder is required to decode the sequences, limiting the number of states in the decoder trellis. The resulting \ac{BCJR}-based decoders can decode and fine-tune codes that have been learned as a turbo-autoencoder or even learn neural codes from scratch.

\vspace{-0.15cm}

\bibliographystyle{IEEEtran}
\bibliography{IEEEabrv,references}

\end{document}

%% file: macros.tex
\renewcommand{\vec}[1]{\mathbf{#1}}

\newcommand{\cv}{\vec{c}}

\newcommand{\ev}{\vec{e}}

\newcommand{\nv}{\vec{n}}

\newcommand{\uv}{\vec{u}}
\newcommand{\vv}{\vec{v}}

\newcommand{\xv}{\vec{x}}
\newcommand{\yv}{\vec{y}}

\newcommand{\Lm}{\vec{L}}

\newcommand{\RR}{\mathbb{R}}

%% file: acronyms.tex
\begin{acronym}
 \acro{CSI}{channel state information}
 \acro{UE}{user equipment}
 \acro{UL}{uplink}
 \acro{BS}{basestation}
 \acro{TDD}{time division duplex}
 \acro{FDD}{frequency division duplex}
 \acro{ECC}{error-correcting code}
 \acro{MLD}{maximum likelihood decoding}
 \acro{HDD}{hard decision decoding}
 \acro{IF}{intermediate frequency}
 \acro{RF}{radio frequency}
 \acro{SDD}{soft decision decoding}
 \acro{NND}{neural network decoding}
 \acro{CNN}{convolutional neural network}
 \acro{ML}{maximum likelihood}
 \acro{GPU}{graphical processing unit}
 \acro{BP}{belief propagation}
 \acro{LTE}{Long Term Evolution}
 \acro{BER}{bit error rate}
 \acro{DER}{detection error rate}
 \acro{SNR}{signal-to-noise-ratio}
 \acro{ReLU}{rectified linear unit}
 \acro{BPSK}{binary phase shift keying}
 \acro{QPSK}{quadrature phase shift keying}
 \acro{AWGN}{additive white Gaussian noise}
 \acro{MSE}{mean squared error}
 \acro{LLR}{log-likelihood ratio}
 \acro{MAP}{maximum a posteriori}
 \acro{NVE}{normalized validation error}
 \acro{BCE}{binary cross-entropy}
 \acro{CE}{cross-entropy}
 \acro{BLER}{block error rate}
 \acro{SQR}{signal-to-quantisation-noise-ratio}
 \acro{MIMO}{multiple-input multiple-output}
 \acro{OFDM}{orthogonal frequency division multiplex}
 \acro{RF}{radio frequency}
 \acro{LOS}{line of sight}
 \acro{NLoS}{non-line of sight}
 \acro{NMSE}{normalized mean squared error}
 \acro{CFO}{carrier frequency offset}
 \acro{SFO}{sampling frequency offset}
 \acro{IPS}{indoor positioning system}
 \acro{TRIPS}{time-reversal IPS}
 \acro{RSSI}{received signal strength indicator}
 \acro{MIMO}{multiple-input multiple-output}
 \acro{ENoB}{effective number of bits}
 \acro{AGC}{automated gain control}
 \acro{ADC}{analog to digital converter}
 \acro{ADCs}{analog to digital converters}
 \acro{FB}{front bandpass}
 \acro{FPGA}{field programmable gate array}
 \acro{JSDM}{Joint Spatial Division and Multiplexing}
 \acro{NN}{neural network}
 \acro{IF}{intermediate frequency}
 \acro{LoS}{line-of-sight}
 \acro{NLoS}{non-line-of-sight}
 \acro{DSP}{digital signal processing}
 \acro{AFE}{analog front end}
 \acro{SQNR}{signal-to-quantisation-noise-ratio}
 \acro{SINR}{signal-to-interference-noise-ratio}
 \acro{ENoB}{effective number of bits}
 \acro{AGC}{automated gain control}
 \acro{PCB}{printed circuit board}
 \acro{EVM}{error vector mangnitude}
 \acro{CDF}{cumulative distribution function}
 \acro{MRC}{maximum ratio combining}
 \acro{MRP}{maximum ratio precoding}
 \acro{MRT}{maximum ratio transmission}
 \acro{DeepL}{deep-learning}
 \acro{DL}{deep learning}
 \acro{SISO}{single-input single-output}
 \acro{SGD}{stochastic gradient descent}
 \acro{CP}{cyclic prefix}
 \acro{MISO}{Multiple Input Single Output}
 \acro{LMMSE}{linear minimum mean square error}
 \acro{ZF}{zero forcing}
 \acro{USRP}{universal software radio peripheral}
 \acro{RNN}{recurrent neural network}
 \acro{GRU}{gated recurrent unit}
 \acro{LSTM}{long short-term memory}
 \acro{NTM}{neural turing machine}
 \acro{DNC}{differentiable neural computer}
 \acro{TCN}{temporal convolutional network}
 \acro{FCL}{fully connected layer}
 \acro{MANN}{memory augmented neural network}
 \acro{RNN}{recurrent neural network}
 \acro{DNN}{dense neural network}
 \acro{FIR}{finite impulse response}
 \acro{BPTT}{back-propagation through time}
 \acro{GAN}{generative adversarial network}
 \acro{ELU}{exponential linear unit}
 \acro{tanh}{hyperbolic tangent}
 \acro{BICM}{bit-interleaved coded modulation}
 \acro{OTA}{over-the-air}
 \acro{IM}{intensity modulation}
 \acro{DD}{direct detection}
 \acro{RL}{reinforcement learning}
 \acro{SDR}{software-defined radio}
 \acro{WGAN}{Wasserstein generative adversarial network}
 \acro{BMD}{bit-metric decoding}
 \acro{BMI}{bit-wise mutual information}
 \acro{LDPC}{low-density parity-check}
 \acro{IDD}{iterative demapping and decoding}
 \acro{IEDD}{iterative equalization, demapping and decoding}
 \acro{JSD}{Jensen-Shannon divergence}
 \acro{MMSE}{minimum mean square error}
 \acro{FFT}{fast Fourier transform}
 \acro{IFFT}{inverse fast Fourier transform}
 \acro{QAM}{quadrature amplitude modulation}
 \acro{EMD}{earth mover's distance}
 \acro{TDL}{tapped delay line}
 \acro{KL}{Kullback-Leibler}
 \acro{PRACH}{physical random access channel}
 \acro{URLLC}{ultra-reliable low-latency communication}
 \acro{ANOMA}{asynchronous non-orthogonal multiple access}
 \acro{FEC}{forward error correction}
 \acro{NOMA}{non-orthogonal medium access}
 \acro{MTC}{machine-type communications}
 \acro{mMTC}{massive machine-type communications}
 \acro{MCS}{modulation and coding scheme}
 \acro{PAPR}{peak-to-average power ratio}
 \acro{MAC}{medium access control}
 \acro{STO}{sampling time offset}
 \acro{STE}{straight-through estimator}
 \acro{PHY}{physical}
 \acro{CCE}{categorical cross-entropy}
 \acro{IoT}{Internet of Things}
 \acro{CCDF}{complementary cumulative distribution function}
 \acro{CRC}{cyclic redundancy check}
 \acro{ACLR}{adjacent channel leakage ratio}
\acro{MD}{missed detection}
 \acro{FA}{false alarm}
 \acro{FAR}{false alarm rate}
 \acro{BCJR}{Bahl-Cocke-Jelinek-Raviv}
 \acro{SCNC}{serially concatenated neural code}

\end{acronym}

%% file: corporateColours.tex
\definecolor{mittelblau}{RGB}{0, 126, 198}
\definecolor{violettblau}{cmyk}{0.9, 0.6, 0, 0}
\definecolor{rot}{RGB}{238, 28 35}
\definecolor{apfelgruen}{RGB}{140, 198, 62}
\definecolor{gelb}{RGB}{1, 221, 0}
\definecolor{orange}{RGB}{244, 111, 33}
\definecolor{pink}{RGB}{237, 0, 140}
\definecolor{lila}{RGB}{128, 10, 145}
\definecolor{hellgrau}{RGB}{224, 224, 224}
\definecolor{mittelgrau}{RGB}{128, 128, 128}
\definecolor{dunkelgrau}{RGB}{80,80,80}
\definecolor{anthrazit}{RGB}{19, 31, 31}

%% file: plotcyclelists.tex
\pgfplotscreateplotcyclelist{corporate colours markers}{%
rot, every mark/.append style={fill=.!80!rot},mark=*\\%
mittelblau, every mark/.append style={fill=.!80!mittelblau},mark=square*\\%
apfelgruen, every mark/.append style={fill=.!80!apfelgruen},mark=triangle*\\%
orange, mark=star\\%
pink, every mark/.append style={fill=.!80!pink},mark=diamond*\\%
violettblau, every mark/.append style={fill=.!80!violettblau},mark=otimes*\\%
lila, mark=|\\%
gelb, every mark/.append style={fill=.!80!gelb},mark=pentagon*\\%
hellgrau, mark=text,text mark=p\\%
anthrazit, mark=text,text mark=a\\%
}

\pgfplotscreateplotcyclelist{corporate colours markers double}{%
rot, every mark/.append style={fill=.!80!rot},mark=*\\%
rot, dashed, every mark/.append style={fill=.!80!rot,solid},mark=*\\%
mittelblau, every mark/.append style={fill=.!80!mittelblau},mark=square*\\%
mittelblau, dashed, every mark/.append style={fill=.!80!mittelblau,solid},mark=square*\\%
apfelgruen, every mark/.append style={fill=.!80!apfelgruen},mark=triangle*\\%
apfelgruen, dashed, every mark/.append style={fill=.!80!apfelgruen,solid},mark=triangle*\\%
orange, mark=star\\%
orange, dashed, mark=star\\%
pink, every mark/.append style={fill=.!80!pink},mark=diamond*\\%
pink, dashed, every mark/.append style={fill=.!80!pink,solid},mark=diamond*\\%
violettblau, every mark/.append style={fill=.!80!violettblau},mark=otimes*\\%
violettblau, dashed, every mark/.append style={fill=.!80!violettblau,solid},mark=otimes*\\%
lila, mark=|\\%
lila, dashed, mark=|\\%
gelb, every mark/.append style={fill=.!80!gelb},mark=pentagon*\\%
gelb, dashed, every mark/.append style={fill=.!80!gelb,solid},mark=pentagon*\\%
}

\pgfplotsset{
  colormap/magma/.style={%
    /pgfplots/colormap={magma}{%
      rgb=(0.001462, 0.000466, 0.013866)
      rgb=(0.035520, 0.028397, 0.125209)
      rgb=(0.102815, 0.063010, 0.257854)
      rgb=(0.191460, 0.064818, 0.396152)
      rgb=(0.291366, 0.064553, 0.475462)
      rgb=(0.384299, 0.097855, 0.501002)
      rgb=(0.475780, 0.134577, 0.507921)
      rgb=(0.569172, 0.167454, 0.504105)
      rgb=(0.664915, 0.198075, 0.488836)
      rgb=(0.761077, 0.231214, 0.460162)
      rgb=(0.852126, 0.276106, 0.418573)
      rgb=(0.925937, 0.346844, 0.374959)
      rgb=(0.969680, 0.446936, 0.360311)
      rgb=(0.989363, 0.557873, 0.391671)
      rgb=(0.996580, 0.668256, 0.456192)
      rgb=(0.996727, 0.776795, 0.541039)
      rgb=(0.992440, 0.884330, 0.640099)
      rgb=(0.987053, 0.991438, 0.749504)
    },
  },
  colormap/inferno/.style={%
    /pgfplots/colormap={inferno}{%
      rgb=(0.001462, 0.000466, 0.013866)
      rgb=(0.037668, 0.025921, 0.132232)
      rgb=(0.116656, 0.047574, 0.272321)
      rgb=(0.217949, 0.036615, 0.383522)
      rgb=(0.316282, 0.053490, 0.425116)
      rgb=(0.410113, 0.087896, 0.433098)
      rgb=(0.503493, 0.121575, 0.423356)
      rgb=(0.596940, 0.154848, 0.398125)
      rgb=(0.688653, 0.192239, 0.357603)
      rgb=(0.775059, 0.239667, 0.303526)
      rgb=(0.851384, 0.302260, 0.239636)
      rgb=(0.912966, 0.381636, 0.169755)
      rgb=(0.956852, 0.475356, 0.094695)
      rgb=(0.981895, 0.579392, 0.026250)
      rgb=(0.987464, 0.690366, 0.079990)
      rgb=(0.973088, 0.805409, 0.216877)
      rgb=(0.947594, 0.917399, 0.410665)
      rgb=(0.988362, 0.998364, 0.644924)
    },
  },
  colormap/plasma/.style={%
    /pgfplots/colormap={plasma}{%
      rgb=(0.050383, 0.029803, 0.527975)
      rgb=(0.186213, 0.018803, 0.587228)
      rgb=(0.287076, 0.010855, 0.627295)
      rgb=(0.381047, 0.001814, 0.653068)
      rgb=(0.471457, 0.005678, 0.659897)
      rgb=(0.557243, 0.047331, 0.643443)
      rgb=(0.636008, 0.112092, 0.605205)
      rgb=(0.706178, 0.178437, 0.553657)
      rgb=(0.768090, 0.244817, 0.498465)
      rgb=(0.823132, 0.311261, 0.444806)
      rgb=(0.872303, 0.378774, 0.393355)
      rgb=(0.915471, 0.448807, 0.342890)
      rgb=(0.951344, 0.522850, 0.292275)
      rgb=(0.977856, 0.602051, 0.241387)
      rgb=(0.992541, 0.687030, 0.192170)
      rgb=(0.992505, 0.777967, 0.152855)
      rgb=(0.974443, 0.874622, 0.144061)
      rgb=(0.940015, 0.975158, 0.131326)
    },
  },
}

%% file: tikz/cnn.tex
\begin{tikzpicture}[
       start chain = 1 going below,    node distance = 0pt,
cell/.style={draw, fill=white, minimum width=2em, minimum height=2em, 
                outer sep=0pt, on chain},
  ]
  
\node [cell] (u0) at (0,4) {};
\foreach \x in {1,...,10} {
    \node [cell] (u\x) {};
}

\node [cell, right=3.4cm of u0, yshift=.4cm] (vg0) {};
\foreach \x in {1,...,10} {
    \node [cell] (vg\x) {};
}

\node [cell, right=3.2cm of u0, yshift=.2cm] (vf0) {};
\foreach \x in {1,...,10} {
    \node [cell] (vf\x) {};
}

\node [cell, right=3cm of u0] (v0) {};
\foreach \x in {1,...,10} {
    \node [cell] (v\x) {};
}

\node [cell, right=2.4cm of v0, yshift=.4cm] (xg0) {};
\foreach \x in {1,...,10} {
    \node [cell] (xg\x) {};
}

\node [cell, right=2.2cm of v0, yshift=.2cm] (xf0) {};
\foreach \x in {1,...,10} {
    \node [cell] (xf\x) {};
}

\node [cell, right=2cm of v0] (x0) {};
\foreach \x in {1,...,10} {
    \node [cell] (x\x) {};
}

\draw [dashed] (x5.north west) to (v3.north east);
\draw [dashed] (x5.south west) to (v7.south east);

\draw [dashed]  (v3.north west) to ($(v3.north west)!0.30!(u1.north east)$); 
\draw [loosely dotted]  ($(v3.north west)!0.30!(u1.north east)$) to ($(v3.north west)!0.70!(u1.north east)$); 
\draw [dashed] ($(v3.north west)!0.70!(u1.north east)$) to (u1.north east); 

\draw [dashed]  (v7.south west) to ($(v7.south west)!0.30!(u9.south east)$); 
\draw [loosely dotted]  ($(v7.south west)!0.30!(u9.south east)$) to ($(v7.south west)!0.70!(u9.south east)$); 
\draw [dashed] ($(v7.south west)!0.70!(u9.south east)$) to (u9.south east); 

\node [] at (x5) {$i$};
\node [xshift=-1cm] at (v3) {$i-\frac{\kappa-1}{2}$};
\node [xshift=-1cm] at (v7) {$i+\frac{\kappa-1}{2}$};

\node [] at ($(u5)!0.5!(v5)$) {$\cdots$};

\node [above=0.3cm of u0, text width=3cm, text centered] {Input Layer\\$\mathbf{v}^{(0)}$};
\node [above=0.3cm of v0, text width=3cm, text centered] {Layer $L-1$\\$\mathbf{v}^{(L-1)}$};
\node [above=0.3cm of x0, text width=3cm, text centered] {Layer $L$\\$\mathbf{v}^{(L)}$};

\draw[<->] ([shift={(-0.7cm,0)}] u9.south) -- ([shift={(-0.7cm,0)}] u1.north)  node [midway,sloped, yshift=0.3cm] {receptive field $R=L(\kappa-1)+1$};

\draw[<-] ([shift={(0.4cm,-0.3)}] x10.south) -- ([shift={(.9cm,0.2)}] xg10.south)  node [midway,sloped, yshift=-0.3cm] {depth};

\draw[->] ([shift={(.9cm,0.2)}] xg10.south) -- ([shift={(.9cm,2.0)}] xg10.south)  node [midway,sloped, yshift=-0.3cm] {position};

\end{tikzpicture}

%% file: tikz/serial_encoder.tex
\begin{tikzpicture}

\tikzstyle{dots}=[line width=1pt , line cap=round,dash pattern=on 0pt off 4pt,black]

\tikzstyle{arrow} = [thick,->, rounded corners=5pt]

\tikzstyle{loosely dotted}=          [dash pattern=on \pgflinewidth off 8pt]

\tikzstyle{NN} =[rectangle, rounded corners,   align=flush center,minimum width=3cm,minimum height=1.0cm,text centered,
	draw=red, fill=red!0 , thick , label={[red]:NN}];
	
\tikzstyle{box} =[rectangle,   align=flush center,minimum width=1.0cm, minimum height=1.1cm,text centered,
	draw=black, fill=red!0 , thick, text=black];
\tikzstyle{box1} =[rectangle,  dashed,  align=flush center,minimum width=4.5cm, minimum height=4.5cm,text centered,
	draw=blue, fill=red!0 , thick , label={[black]:}];	
\tikzstyle{boxText} =[rectangle,  align=flush center,minimum width=15.74776cm,
	minimum height=15.74776cm,text centered,
	draw=black!50, fill=red!0 , thick ];
\tikzstyle{interleaver} =[rectangle,   align=flush center,minimum width=1.0cm, minimum height=1.0cm,text centered,
	draw=black, fill=red!0 , thick, text=black];
\tikzstyle{plus} =[circle,draw=black, fill=white, inner sep=2pt,minimum size=5pt];

\node [box1, draw=apfelgruen, minimum width=2.7cm, minimum height=1.8cm, label={[black]:Outer encoder $f_\mathrm{O}(\cdot)$}] at (0.7,4.1) {};

\node [box1, draw=apfelgruen, minimum width=2.75cm, minimum height=1.8cm, label={[black]:Inner encoder $f_\mathrm{I}(\cdot)$}] at (6.11,4.1) {};

	\node [] (start) at (-1.75,4) {};	
	\node [box,label={[black]:}] (enc1) at (0,4) {Enc O\\CNN};
	\node [interleaver] (sign) at (1.5,4) {Bina-\\rizer};
	\node [interleaver] (int) at (3.85,4) {$\vec{\pi}$};
	\node [box,label={[black]:}] (enc2) at (5.35,4) {Enc I\\ CNN};
	\node [box,label={[black]:}] (norm) at (6.85,4) {Norm};

	\draw [arrow] (start)-- node[midway, above]{$\uv \in $}node[midway, below]{$ \mathbb{F}_2^{k } $} (enc1);
	\draw [arrow,double] (enc1) -- node[midway, above]{$\cv_{\RR}$}node[midway, below]{} (sign);
	\draw [arrow,double] (sign) -- node[midway, above]{$\cv \in $}  node[midway, below]{\footnotesize $\{ \pm 1 \}^{k \times F_{\cv}}$} (int);
	\draw [arrow,double] (int) -- node[midway, above]{$\cv_{\pi}$} (enc2);
	\draw [arrow] (enc2) -- node[midway, above]{} (norm);
	\draw [arrow] (norm) -- (8.,4)  node[midway, above]{$\xv$} node[midway, below, xshift=.2cm]{$\in \mathbb{R}^{n} $};

\end{tikzpicture}

%% file: tikz/serial_decoder.tex
\begin{tikzpicture}

\tikzstyle{dots}=[line width=1pt , line cap=round,dash pattern=on 0pt off 4pt,black]

\tikzstyle{arrow} = [thick,->]

\tikzstyle{loosely dotted}=          [dash pattern=on \pgflinewidth off 8pt]

\tikzstyle{NN} =[rectangle, rounded corners,   align=flush center,minimum width=3cm,minimum height=1.0cm,text centered,
	draw=red, fill=red!0 , thick , label={[red]:NN}];
	
\tikzstyle{box} =[rectangle,   align=flush center,minimum width=1.0cm, minimum height=1.1cm,text centered,
	draw=black, fill=red!0 , thick, text=black];
\tikzstyle{box1} =[rectangle, dashed,  align=flush center,minimum width=4.5cm, minimum height=4.5cm,text centered,
	draw=blue, fill=red!0 , thick , label={[black]:}];	
\tikzstyle{boxText} =[rectangle,  align=flush center,minimum width=15.74776cm,
	minimum height=15.74776cm,text centered,
	draw=black!50, fill=red!0 , thick ];
\tikzstyle{interleaver} =[rectangle,  align=flush center,minimum width=1.0cm, minimum height=1.0cm,text centered,
	draw=black, fill=red!0 , thick, text=black];
\tikzstyle{plus} =[circle,draw=black, fill=white, inner sep=2pt,minimum size=5pt];

\node [box1, draw=mittelblau, minimum width=5.33cm, minimum height=3.5cm, label={[black]:Decoder}] at (2.02,3.4) {};
	
	\node [] (start) at (-1.7,4) {};	
	\node [box,label={[black]:Inner}] (dec1) at (0,4) {Dec I\\ CNN/ \\ BCJR};
	\node [box,label={[black]:Outer}] (dec2) at (4,4) {Dec O\\ CNN/ \\ BCJR};
	\node [interleaver] (deint) at (2,4) {$\vec{\pi}^{-1}$};
	\node [interleaver] (int) at (2,2.5) {$\vec{\pi}$};
	
	\draw [arrow] (start)-- node[pos=0.5, above]{$\vec{y}$} (dec1);
	\draw [arrow,double] (dec1) -- node[pos=0.5, below]{$\Lm_{\cv,\pi}^\mathrm{E}$} (deint);
	\draw [arrow,double] (deint) -- node[pos=0.5, below]{$\Lm_{\cv}^{\mathrm{E}}$} (dec2);

	\draw [arrow] (dec2) --node[pos=0.6, above]{$\Lm_{\uv}^\mathrm{T}$}node[pos=0.6, below]{$\in \RR^{k} $} (5.5,4);
	\draw [arrow,double] (dec2) |-node[pos=0.65, above]{$\Lm_{\cv}^{'\mathrm{E}}$}node[pos=0.65, below]{$\in \RR^{k \times F} $} (int);
	\draw [arrow,double] (int) -| node[pos=0.5, near start, below]{$\Lm_{\cv,\pi}^{'\mathrm{E}}$} (dec1);

\end{tikzpicture}

%% file: tikz/effective_encoder.tikz
\begin{tikzpicture}

\tikzstyle{dots}=[line width=1pt , line cap=round,dash pattern=on 0pt off 4pt,black]

\tikzstyle{arrow} = [thick,->]

\tikzstyle{loosely dotted}=          [dash pattern=on \pgflinewidth off 8pt]

\tikzstyle{NN} =[rectangle, rounded corners,   align=flush center,minimum width=3cm,minimum height=1.0cm,text centered,
	draw=red, fill=red!0 , thick , label={[red]:NN}];
	
\tikzstyle{box} =[rectangle,   align=flush center,minimum width=1.0cm, minimum height=1.0cm,text centered,
	draw=black, fill=red!0 , thick, text=black];
\tikzstyle{box1} =[rectangle,  dashed,  align=flush center,minimum width=4.5cm, minimum height=4.5cm,text centered,
	draw=blue, fill=red!0 , thick , label={[black]:}];	
\tikzstyle{boxText} =[rectangle,  align=flush center,minimum width=15.74776cm,
	minimum height=15.74776cm,text centered,
	draw=black!50, fill=red!0 , thick ];
\tikzstyle{interleaver} =[rectangle,   align=flush center,minimum width=1.0cm, minimum height=1.0cm,text centered,
	draw=black, fill=red!0 , thick, text=black];
\tikzstyle{plus} =[circle,draw=black, fill=white, inner sep=2pt,minimum size=5pt];

\node [box1, draw=apfelgruen, minimum width=1.3cm, minimum height=3.3cm, label={[black]:Outer encoder $f_\mathrm{O}(\cdot)$}] at (0.0,4.) {};

\node [box1, draw=apfelgruen, minimum width=4.3cm, minimum height=3.3cm, label={[black]:Inner encoder $f_\mathrm{I}(\cdot)$}] at (5.0,4.0) {};

	\node [] (beforestart) at (-2.55,4) {};	
	\node [] (start) at (-1.75,4) {};	
	\node [box,label={[black]:}] (enc1) at (0,3.15) {$f_\mathrm{O}^{(0)}$};
	\node [box,label={[black]:}] (enc11) at (0,4.85) {$f_\mathrm{O}^{(1)}$};
	\node [interleaver] (int) at (1.75,3.15) {$\vec{\pi}$};
	\node [interleaver] (int1) at (1.75,4.85) {$\vec{\pi}$};
	\node [box,label={[black]:}] (enc2) at (3.5,3.15) {$f_\mathrm{I}^{(0)}$};
	\node [box,label={[black]:}] (enc21) at (3.5,4.85) {$f_\mathrm{I}^{(1)}$};
	\node [box,label={[black]:}] (mux) at (4.75,4) {MUX};
	\node [box,label={[black]:}] (norm) at (6.5,4) {Norm};

	\draw [arrow] (beforestart)-- node[midway, above]{$\uv $} node[midway, below]{$\in \mathbb{F}_2^{k } $} (-1.25,4) |-  (enc1);
	\draw [arrow] (beforestart)-- node[midway, above]{$\uv $} node[midway, below]{$\in \mathbb{F}_2^{k } $} (-1.25,4) |-  (enc11);
    
    \draw [arrow] (enc1) -- node[midway, above]{$\cv_0 $} (int);
	\draw [arrow] (enc11) --  node[midway, above]{$\cv_1 $}(int1);
	
	\draw [arrow] (int) -- (enc2);
	\draw [arrow] (int1) --  (enc21);

	\draw [arrow] (enc21) -| node[midway, above]{$\xv_1 $}(mux);
	\draw [arrow] (enc2) -| node[midway, below]{$\xv_0 $} (mux);
	
	\draw [arrow] (mux) -- node[midway, above]{} (norm);
	\draw [arrow] (norm) -- (8.,4)  node[midway, above]{$\xv$} node[midway, below, xshift=.2cm]{$\in \mathbb{R}^{n} $};

\end{tikzpicture}

%% file: tikz/gradient_energy_outer.tikz
\begin{tikzpicture}
\begin{axis}[
    width=\linewidth,
	    height=0.55\linewidth,
        clip mode=individual,
        xmajorgrids,
        xminorgrids=true,
        yminorticks=true,
        ymajorgrids,
        yminorgrids,
		xmin=-5,
		xmax=5,
		legend cell align={left},
		ymin=0,
		ymax=1.1e-0,
        ymode=normal,
        legend style={at={(0.0,1.0)},anchor=south west, draw=white!15!black, font=\tiny, /tikz/every even column/.append style={column sep=0.985cm}},
        legend columns=2,
        ylabel={\footnotesize Expected error $e_{\partial},~e_{\Delta}$},
        xlabel={\footnotesize  Omitting input position $j$},
        mark size=2.5pt,
        line width = 1.0pt,
    ]

\addplot+ [color=mittelblau,line width = 0.9pt ,mark=o, mark options={color=mittelblau}] %
table[col sep=comma]{
-10.0,4.16006787e-06
-9.0,1.90559604e-05
-8.0,8.28550765e-05
-7.0,0.000460337993
-6.0,0.000628473528
-5.0,0.00478149112
-4.0,0.0213217381
-3.0,0.099711746
-2.0,1.0
-1.0,0.244975492
0.0,0.848592222
1.0,0.663875997
2.0,0.0952512026
3.0,0.0776395947
4.0,0.0110821845
5.0,0.00285771373
6.0,0.000620238774
7.0,0.000123684731
8.0,3.74367737e-05
9.0,2.22118251e-06
10.0,3.14024589e-07

};
\label{plot:beta1.18}
\addlegendentry{\footnotesize $f_{\mathrm{O}}^{(0)}$ approximation};

\addplot+ [color=rot,line width = 0.9pt ,mark=o, mark options={color=rot,fill=rot, solid}] %
table[col sep=comma]{
-10.0,5.8631454e-06
-9.0,1.2400951e-05
-8.0,6.072364e-05
-7.0,0.000202838
-6.0,0.00063251139
-5.0,0.0028542946
-4.0,0.015326637
-3.0,0.047101565
-2.0,0.13443284
-1.0,0.93512326
0.0,0.67421699
1.0,0.38545021
2.0,0.16901696
3.0,1.0
4.0,0.026599212
5.0,0.0055425712
6.0,0.001107856
7.0,0.00024353755
8.0,3.5869169e-05
9.0,6.7624355e-06
10.0,5.4589265e-07

};
\label{plot:beta1.18}
\addlegendentry{\footnotesize $f_{\mathrm{O}}^{(1)}$ approximation};

\addplot+ [color=mittelblau,line width = 0.9pt ,dashed,mark=triangle,mark options={color=mittelblau,fill=mittelblau, solid}] %
table[col sep=comma]{
-3.0,0.0
-2.0,1.0
-1.0,0.0
0.0,1.0
1.0,1.0
2.0,0.0
3.0,0.0

};
\label{plot:beta1.18}
\addlegendentry{\footnotesize $f_{\mathrm{O}}^{(0)}$ verification};

\addplot+ [color=rot,line width = 0.9pt,mark=triangle ,dashed, mark options={color=rot,fill=rot, solid}] %
table[col sep=comma]{
-3.0,0.0
-2.0,0.0
-1.0,1.0
0.0,1.0
1.0,0.0
2.0,0.0
3.0,1.0

};
\label{plot:beta1.18}
\addlegendentry{\footnotesize $f_{\mathrm{O}}^{(1)}$ verification};

\end{axis}
\end{tikzpicture}

%% file: tikz/gradient_energy.tikz
\begin{tikzpicture}
\pgfplotsset{compat=1.5}

\begin{axis}[
    width=\linewidth,
	    height=0.6\linewidth,
        clip mode=individual,
        xmajorgrids,
        xminorgrids=true,
        yminorticks=true,
        ymajorgrids,
        yminorgrids,
		xmin=-6,
		xmax=6,
		ymin=1e-3,
		ymax=1.2e-0,
        ymode=log,
        legend style={at={(0.0,1.0)},anchor=south west, draw=white!15!black, font=\tiny, /tikz/every even column/.append style={column sep=0.975cm}},
        legend columns=2,
        ylabel={\footnotesize Expected error $e_{\partial},~e_{\Delta}$},
        xlabel={\footnotesize  Omitting input position $j$},
        mark size=2.5pt,
        legend cell align={left},
        cycle list name=corporate colours markers
    ]
  
 \draw[draw=none,fill=white,fill opacity=0.7] (axis cs:3.2,  3.9e-02) rectangle (axis cs:6, 2.1e-02);
 
\draw[ line width=1.0pt, color=black,dashed] (axis cs:-6,  2e-2) -- node[above,pos=0.9]{\footnotesize $t=2\cdot10^{-2}$} (axis cs:6, 2e-2);

\addplot+ [color=mittelblau , mark=o,line width = 1pt, mark options={fill=mittelblau, solid}] %
table[col sep=comma]{
-10.0,4.9462173e-08
-9.0,6.2939125e-07
-8.0,6.285723e-06
-7.0,4.3056993e-05
-6.0,0.00042374447
-5.0,0.0025845792
-4.0,0.021748485
-3.0,0.09048526
-2.0,0.33611894
-1.0,1.0
0.0,0.70218122
1.0,0.38959116
2.0,0.060256626
3.0,0.020147432
4.0,0.0029704669
5.0,0.00067325833
6.0,9.9690464e-05
7.0,1.2510626e-05
8.0,1.1561151e-06
9.0,2.2106735e-07
10.0,7.3157254e-09

};
\label{plot:f0approx}
\addlegendentry{\footnotesize $f_{\mathrm{I}}^{(0)}$ approximation};

\addplot+ [color=rot, mark=o ,line width = 1pt, mark options={fill=rot, solid}] 
table[col sep=comma]{
-10.0,2.2812319e-08
-9.0,1.6721469e-07
-8.0,8.8639268e-07
-7.0,3.2133751e-06
-6.0,2.5449086e-05
-5.0,0.00010024598
-4.0,0.00040658537
-3.0,0.0013486968
-2.0,0.0030471054
-1.0,0.0078407172
0.0,0.038350139
1.0,0.030662645
2.0,1.0
3.0,0.010743189
4.0,0.0098161595
5.0,0.0010536143
6.0,0.00039493761
7.0,4.1790379e-05
8.0,1.5833986e-05
9.0,1.1670811e-06
10.0,1.276333e-07
};
\label{plot:f1approx}
\addlegendentry{\footnotesize $f_{\mathrm{I}}^{(1)}$ approximation};

\addplot+ [color=mittelblau , mark=triangle,dashed,line width = 1pt, mark options={fill=mittelblau, solid}] %
table[col sep=comma]{
-8.0,3.2269066144926556e-06
-7.0,2.9580235723173824e-05
-6.0,0.00036241910264130924
-5.0,0.0034034295956839366
-4.0,0.02158296519140867
-3.0,0.14320468395374228
-2.0,0.4036236251772713
-1.0,1.0
0.0,0.826081343918319
1.0,0.44174609692694244
2.0,0.0592861476050302
3.0,0.029592967772319335
4.0,0.00234940519321797
5.0,0.00041201799719558775
6.0,5.537273760120761e-05
7.0,5.928253787774643e-06
8.0,8.418940780073595e-07

};
\label{plot:f0veri}
\addlegendentry{\footnotesize $f_{\mathrm{I}}^{(0)}$ verification};

\addplot+ [color=rot , mark=triangle,dashed,line width = 1pt, mark options={fill=rot, solid}] 
table[col sep=comma]{
-6.0,5.838480539092923e-06
-5.0,8.74884821148585e-06
-4.0,1.6564031549912384e-05
-3.0,4.302371150307345e-05
-2.0,0.00033741845286733437
-1.0,4.4963451091499674e-05
0.0,0.002857568144019813
1.0,0.0023153515128285273
2.0,1.0
3.0,0.0033707517373005058
4.0,0.0027511622346004833
5.0,0.00017889760048085386
6.0,2.5213718579601378e-05
};
\label{plot:f1veri}
\addlegendentry{\footnotesize $f_{\mathrm{I}}^{(1)}$ verification};

\draw[draw=none,fill=white,fill opacity=0.6] (axis cs:-6.0,  1.9e-02) rectangle (axis cs:6, 1.0e-03);

\draw[ line width=1.0pt, color=mittelblau, <->, draw=mittelblau!80!black] (axis cs:-6,  1.2e-3) -- node[above,xshift=0.2cm]{\footnotesize $\bar{\mathcal{J}}$} (axis cs:-4, 1.2e-3);

\draw[ line width=1.0pt, color=mittelblau, <->, draw=mittelblau!80!black] (axis cs:-4,  1.2e-3) -- node[above,xshift=-0.2cm]{\footnotesize $\mathcal{J}$} (axis cs:3, 1.2e-3);

\draw[ line width=1.0pt, color=mittelblau, <->, draw=mittelblau!80!black] (axis cs:3,  1.2e-3) -- node[above,xshift=0.5cm]{\footnotesize $\bar{\mathcal{J}}$} (axis cs:6, 1.2e-3);

\draw[ line width=0.8pt, color=black,dotted] (axis cs:-4,  1.2e-3) -- (axis cs:-4, 2e-2);
\draw[ line width=0.8pt, color=black,dotted] (axis cs:3,  2e-2) -- (axis cs:3, 1.2e-3);

\draw[ line width=0.8pt, color=black,dotted] (axis cs:0,  3e-2) -- (axis cs:0, 5e-3);
\draw[ line width=0.8pt, color=black,dotted] (axis cs:2,  1e-0) -- (axis cs:2, 5e-3);

\draw[ line width=1.0pt, color=rot, <->, draw=rot!80!black] (axis cs:-6,  5e-3) -- node[above]{\footnotesize $\bar{\mathcal{J}}$} (axis cs:0, 5e-3);

\draw[ line width=1.0pt, color=rot, <->, draw=rot!80!black] (axis cs:0,  5e-3) -- node[above,xshift=-0.1cm]{\footnotesize $\mathcal{J}$} (axis cs:2, 5e-3);

\draw[ line width=1.0pt, color=rot, <->, draw=rot!80!black] (axis cs:2,  5e-3) -- node[above,xshift=0.4cm]{\footnotesize $\bar{\mathcal{J}}$} (axis cs:6, 5e-3);

\end{axis}
\end{tikzpicture}

%% file: tikz/bler_graph.tikz
\begin{tikzpicture}
		
		\pgfplotsset{compat=1.5}

		\begin{axis}[
			xmode=normal,
			ymode=log,
			xlabel=\footnotesize $E_\mathrm{b}/N_0~(\mathrm{dB})$, %
			ylabel=\footnotesize $\mathrm{BLER}$,
			xmin = 2,
			xmax=4.6,
			ymax=10^(-1),
			ymin=10^(-5),
			mark size=2.5pt,
			legend style={at={(axis cs:2,0.00001)},anchor=south west}, 
			grid=both,
			minor grid style={gray!25},
			major grid style={gray!25},
			width=\linewidth,
	        height=0.75\linewidth,
			cycle list name=corporate colours markers,
			legend cell align={left},
			line width=0.8pt, %
			]

			\addplot+ [anthrazit, mark=o, dashed, mark options={ solid}] %
			table[x=snr, y=cer]{./tikz/results/bler/ARA.txt};
    		\label{plot:ldpc_ccsds}
            \addlegendentry{\footnotesize LDPC AR3A \cite{liva2016code}};

			\addplot+ [apfelgruen, mark=square, dashed, mark options={ solid}]
			table[x=snrb_turbo_lte_maxlog_R12,y=bler_turbo_lte_maxlog_R12, ,col sep=comma]{./tikz/results/bler/turbo_lte_maxlog_64_128_4iters.txt};
			\label{plot:turbo_lte_maxlog_64_128}
            \addlegendentry{\footnotesize Turbo LTE \cite{3gpp2007standard}};

			\addplot+ [mittelblau, mark options={fill=mittelblau, solid},mark=square, line width=1.3pt, mark size= 3pt] 
			table[x expr=\thisrowno{0}+0.06 ,y=bler_1p,col sep=comma]{./tikz/results/bler/bcjr_delta11_it6_tb16_maplog.txt};
			\label{plot:bcjr_delta11_it6_tb16}
            \addlegendentry{\footnotesize \ac{SCNC} - BCJR 6 iters};

			\addplot+ [mittelblau, mark options={fill=mittelblau, solid},dotted,mark=o, line width=1.0pt, mark size= 2.5pt] 
			table[x expr=\thisrowno{0}+0.00 ,y=bler,col sep=comma]{./tikz/results/bler/bcjr_delta11_it6_tb16_maplog.txt};
			\label{plot:bcjr_delta11_it6_tb16_no_bf}
            \addlegendentry{\footnotesize \dittoclosing without parity bit};
			
			\addplot+ [rot, mark options={fill=rot, solid},mark=diamond, line width=1.3pt, mark size= 3pt] 
			table[x expr=\thisrowno{0}+0.06 ,y=bler_1p,col sep=comma]{./tikz/results/bler/stae_delta11.txt};
			\label{plot:stae_delta11}
            \addlegendentry{\footnotesize \ac{SCNC} - NN 6 iters};
            
            \addplot+ [mittelblau, mark=triangle,mark options={fill=mittelblau, solid}, line width=1.3pt, mark size= 3pt] 
			table[x expr=\thisrowno{0}+0.06 ,y=bler_1p,col sep=comma]{./tikz/results/bler/bcjr_delta11_tb32_it32_map.txt};
			\label{plot:bcjr_delta11_tb32_it32_map}
            \addlegendentry{\footnotesize \ac{SCNC} - BCJR 32 iters};

            \addplot+ [apfelgruen, mark=o, dashed, mark options={ solid}] 
			table[x=snr, y=cer]{./tikz/results/bler/turbo16.txt};
    		\label{plot:turbo16}
            \addlegendentry{\footnotesize Turbo 16-states \cite{liva2016code}};

		\end{axis}
		
		\begin{axis}[
		    axis x line=none,
		    axis y line=none,
		    ymode=log,
		    xmin = 2,
			xmax= 4.5,
			ymax=10^(-1),
			ymin=10^(-5),
			width=\linewidth,
	        height=0.75\linewidth,
			mark size=2.0pt,
    		legend style={at={(axis cs:4.5,0.1)},anchor=north east},
    		cycle list name=corporate colours markers,
			legend cell align={left},
			line width=0.8pt, %
    		]

		\end{axis}

	\end{tikzpicture}

%% file: tikz/bler_train_graph.tikz
\begin{tikzpicture}
		
		\pgfplotsset{compat=1.5}
		\tikzset{font={\fontsize{8pt}{8}\selectfont}}

		\begin{axis}[
			xmode=normal,
			ymode=log,
			xlabel=$E_\mathrm{b}/N_0~(\mathrm{dB})$, %
			ylabel=$\mathrm{BLER}$,
			xmin = 2,
			xmax=4.56,
			ymax=10^(-1),
			ymin=10^(-4),
			mark size=2.5pt,
			legend style={at={(axis cs:2,0.0001)},anchor=south west}, 
			grid=both,
			minor grid style={gray!25},
			major grid style={gray!25},
			width=\linewidth,
	        height=0.65\linewidth,
			cycle list name=corporate colours markers,
			legend cell align={left},
			line width=1pt, %
			]

			\addplot+ [mittelblau, mark=otimes,mark options={fill=mittelblau, solid}] 
			table[x expr=\thisrowno{0}+0.0 ,y=bler_1p,col sep=comma]{./tikz/results/bler/bcjr_delta11_it2_tb16_logmap_m5_m3_m1.txt};
			\label{plot:bcjr_delta11_it2_tb16_logmap_m5_m3_m1}
            \addlegendentry{\footnotesize SCNC - BCJR 2 iters, $m_\xv^{(0)}=3$};

            \addplot+ [orange, mark options={fill=orange, solid},mark=square] 
			table[x expr=\thisrowno{0}+0.06 ,y=bler_1p,col sep=comma]{./tikz/results/bler/bcjr_delta11_it2_tb16_logmap_m5_m3_m1_finetuned.txt};
			\label{plot:bcjr_delta11_it2_tb16_logmap_m5_m3_m1_finetuned}
            \addlegendentry{\footnotesize  \dittoclosing finetuned};
            
            \addplot+ [orange,mark=triangle, mark options={fill=orange,solid}] 
			table[x expr=\thisrowno{0}+0.06 ,y=bler_1p,col sep=comma]{./tikz/results/bler/bcjr_delta11_it2_tb16_logmap_m5_m4_m4_scratch.txt};
			\label{plot:bcjr_delta11_it2_tb16_logmap_m5_m4_m4_scratch}
            \addlegendentry{\footnotesize \dittoclosing scratch (inner encoder)};

        	\addplot+ [apfelgruen, mark=square, mark options={solid}, dashed]
			table[x=snrb_turbo_lte_maxlog_R12,y=bler_turbo_lte_maxlog_R12, ,col sep=comma]{./tikz/results/bler/turbo_lte_logmap_64_128_2iters.txt};
			\label{plot:turbo_lte_logmap_64_128_2iters}
            \addlegendentry{\footnotesize Turbo LTE 2 iters \cite{3gpp2007standard}};

		\end{axis}

	\end{tikzpicture}